\documentclass[prd,twocolumn,showpacs,preprintnumbers,superscriptaddress,floatfix,nofootinbib]{revtex4-1}
\usepackage{graphicx,,booktabs}
\usepackage{amssymb}
\usepackage{amsmath,txfonts}
\usepackage{graphicx}
\usepackage{dcolumn}
\usepackage{color}
\usepackage{bm}
\usepackage[colorlinks,
            citecolor=blue,
            anchorcolor=green,
            menucolor=orange,
            linkcolor=red,
            filecolor=red,
            runcolor=pink,
            urlcolor=blue,
            frenchlinks=red]{hyperref}

\begin{document}

\title{Production of $\phi(2170)$ and $\eta(2225)$ in a kaon induced reaction}
\author{Xiao-Yun Wang}
\thanks{xywang01@outlook.com}
\affiliation{Department of physics, Lanzhou University of Technology,
Lanzhou 730050, China}
\author{Jun He}
\thanks{Corresponding author : junhe@njnu.edu.cn}
\affiliation{Department of  Physics and Institute of Theoretical Physics, Nanjing Normal University,
Nanjing, Jiangsu 210097, China}
\begin{abstract}
In this work, we study the production of strange quarkoniums, the $\phi(2170)$, also named  $Y(2175)$, and the $\eta(2225)$, via a kaon induced reaction on a proton target in an effective Lagrangian approach. The total and differential cross sections of the reactions $K^{-}p\rightarrow \phi (2170)\Lambda $ and $K^{-}p\rightarrow \eta (2225)\Lambda $ are calculated by the Reggeized $t $-channel Born term  under an assumption that the $\phi(2170)$ and $\eta(2225)$ are $\Lambda\bar{\Lambda}$ molecular states. At the center of mass energies of about 4.2 GeV, the total cross section for the $\phi(2170)$ production is predicted to be about  1 $\mu $b. The numerical results indicate that it is feasible to produce the $\phi (2170)$  via kaon beam scattering at the best energy window near 4.2 GeV.  The total cross section for the $\eta(2225)$ production is smaller than that for the $\phi(2170)$ production and it may reach an order of the magnitude of 0.1 $\mu$b. The differential cross sections for both reactions at  different center of mass energies are also presented. It is found that the Reggeized $t$ channel gives a considerable contribution at forward angles. As the energy increases, the contribution from the  $t$-channel  almost concentrates  at extreme forward angles. From these theoretical predictions, the relevant experimental research is suggested, which could provide important information to clarify the internal structure and production mechanism of these two strange quarkoniums.
\end{abstract}

\maketitle

\section{Introduction}

The $K$ meson was first observed by the physicists Rochester
and Butler~\cite{Rochester:1947mi}, which opened the door to explore
the strange particles. Naturally, the kaon beam becomes a powerful tool to explore strange hadron and hypernucleus.  As of now, many strange quarkoniums have been
observed and were listed in the Review of Particle Physics (PDG)~\cite%
{Tanabashi:2018oca}. However, the internal structure of the strange
quarkoniums is still a confusing problem due to large nonperturbative effect
in the light flavor sector. Some strange quarkoniums were also considered as
exotic states which are beyond the conventional picture of meson composed of
a quark pair. Among these strange quarkoniums, the $\phi (2170)$ and $\eta (2225)$
attract special attention in both experiment and theory.

In the PDG, the $\phi (2170)$, which is also named  $Y(2175)$ in the literature, is listed as a vector state with quantum
numbers $I^{G}(J^{PC})=0^{-}(1^{--})$ with a suggested mass of $2188\pm $
10~MeV and a suggested width of $83\pm 12$~MeV~\cite{Tanabashi:2018oca}. The
existence of the $\eta (2225)$ meson with quantum numbers $%
I^{G}(J^{PC})=0^{+}(1^{-+})$ was confirmed by the BESIII Collaboration ~\cite%
{Ablikim:2016hlu} via a partial wave analysis of the decay process $J/\psi
\rightarrow \gamma \phi \phi $, which has a mass of $2216_{-5-11}^{+4+21}$%
~MeV and a width of $185_{-14-17}^{+12+43}$~MeV.
In conjunction with the experiment activities, the $\phi (2170)$ and $\eta
(2225)$ arouse many concerns from the theoretical side. In the literature, these two states, $%
\phi (2170)$ and $\eta (2225)$, were interpreted as traditional $s\bar{s}$ meson
\cite{Ding:2007pc,Wang:2012wa,Afonin:2014nya,Li:2008we,Li:2008et}, $%
\Lambda \bar{\Lambda}$ bound state \cite%
{Deng:2013aca,Zhao:2013ffn,Dong:2017rmg}, tetraquark state \cite%
{Wang:2006ri,Chen:2008ej,Drenska:2008gr}, hybrid state \cite%
{Ding:2007pc,Ding:2006ya}, or  $\phi KK$ resonance state \cite%
{MartinezTorres:2008gy,GomezAvila:2007ru}.

Since the $\phi (2170)$ and $\eta (2225)$ were observed only in the $J/\psi $
meson decay process or $e^{-}e^{+}$ collision, it is an interesting and
important topic to study these two strange quarkoniums though different
processes. In most theoretical interpretations, there exists a strange quark pair in these two states and they are prone to decay into $K$ and $K^*$ meson. For example, in Ref.~\cite{Dong:2017rmg}, the strong decays of $\phi
(2170) $ and $\eta (2225)$ were calculated by taking the two states as bound
states of $\Lambda \bar{\Lambda}$ in the molecular scenario. One notes that
the partial decay widths of both $\phi (2170)\rightarrow KK$ and $\eta
(2225)\rightarrow K^{\ast }K$ are very large. Thus we expect that the
cross section of these two states produced by kaon beam on proton target will be
large enough to be observed in the existent facilities. The kaon-induced reaction may be an effective way to study
light strange meson, which is available at  J-PARC~\cite{Nagae:2008zz}, JLab~\cite{Amaryan:2017ldw}, 
COMPASS@CERN~\cite{Velghe:2016jjw}, and OKA@U-70~\cite{Obraztsov:2016lhp}. The data
from future experiments at those facilities will provide a good opportunity
to deepen our understanding of internal structure of strange meson.

In this work, the strange quarkonium $\phi (2170)/\eta (2225)$ production
via kaon induced reaction will be investigated. To this end we adopt here an
effective Lagrangian approach in terms of only $t$ channel with $K/K^{\ast
} $ exchange. In general, a phenomenological Regge treatment is successfully applied
in meson production at high energies. The Regge model can reproduce the
energy and $t$ dependence of cross sections especially at  forward angles. Since the experimental data of the  reactions $K^{-}p\rightarrow \phi
(2170)\Lambda $ and $K^{-}p\rightarrow \eta (2225)\Lambda $ are
scare, in this paper, we pay special attention to the energy dependence of
cross sections predicted with Regge model.

This paper is organized as follows. After introduction, we present formalism
including Lagrangians and amplitudes of the $\phi (2170)/\eta (2225)$
production in Section II. Numerical results are discussed in Sec. III,
followed by a brief summary in Sec. IV.

\section{Formalism}

\begin{figure}[t]
\begin{center}
\includegraphics[scale=0.55]{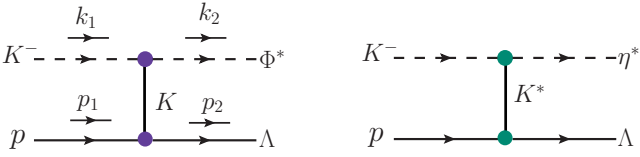}
\end{center}
\caption{(Color online) Feynman diagrams for the $K^{-}p\rightarrow \protect%
\phi (2170)\Lambda $ reaction (left) and for the $K^{-}p\rightarrow \protect%
\eta (2225)\Lambda $ reactions (right).}
\label{Fig: Feynman}
\end{figure}

The strange quarkoniums $\phi (2170)$ and $\eta (2225)$ productions via kaon
induced reactions on a proton target with $t$-channel $K/K^{\ast }$ meson
exchange are depicted in Fig.~\ref{Fig: Feynman}. In Ref.~\cite{Dong:2017rmg}, the partial decay widths of the $\phi (2170)$ and $%
\eta (2225)$ were calculated by considering these two strange quarkoniums as
$\Lambda \bar{\Lambda}$ bound states in the molecular scenario. The results
shown that the partial decay widths $\Gamma _{\phi ^{\ast }\rightarrow
KK}\simeq 73.8-87.7$ MeV and $\Gamma _{\eta ^{\ast }\rightarrow K^{\ast
}K}\simeq 71.1-87.3$ MeV, which are dominant in their total decay widths,
respectively \cite{Dong:2017rmg}. Thus the production of these two strange
quarkoniums will be calculated by taking the $K$ or $K^{\ast }$ exchange as
the dominant contribution in the $t$ channel.

In the present work, the
contribution from $s$ channel  is omitted since it is usually negligibly small~\cite{Wang:2017qcw}. In the current work, the thresholds to produce the  $\phi(2170)$ and the $\eta(2225)$ is higher than 3 GeV. In this energy region, the intermediate $\Lambda^*$  in $s$ channel can be well described by the Reggeized $t$ channel adopted in this work due to the duality. For the $\Lambda$ and $\Lambda^*$   under the threshold, their contributions will be suppressed.  If we assume the couplings between the $\Lambda/\Lambda^*$  and the final state are not unusually strong, the $s$-channel contributions can be neglected. Usually, the contribution of the $u$ channel with $\Lambda$ exchange is also small and negligible at low energies if we assume the relevant couplings are not unusually strong~%
\cite{Wang:2017qcw}. Furthermore, at high energies, the contribution of
the $u$ channel to the total cross section will also become small and
negligible when the Reggeized treatment was applied to the $u$ channel. Hence, the contributions from the  $\Lambda$  in the $u$
channel will not be included in the current calculation.

\subsection{Lagrangians}

For kaon induced production of the $\phi (2170)$ $(\equiv \phi ^{\ast })$ and $%
\eta (2225)$ $(\equiv \eta ^{\ast })$, the relevant Lagrangians for the $t$
channel read as followings \cite{Wan:2015gsl,Ryu:2012tw,Liu:2008qx}%
\begin{eqnarray}
\mathcal{L}_{\phi ^{\ast }KK} &=&-ig_{\phi ^{\ast }KK}[(\partial ^{\mu }K)%
\bar{K}-(\partial ^{\mu }\bar{K})K]\phi _{\mu }^{\ast }, \\
\mathcal{L}_{KN\Lambda } &=&ig_{KN\Lambda }\bar{N}\gamma _{5}\Lambda K{\ +}%
\text{ H.c.} \\
\mathcal{L}_{\eta ^{\ast }K^{\ast }K} &=&ig_{\eta ^{\ast }K^{\ast
}K}(K\partial _{\mu }\eta ^{\ast }-\partial _{\mu }K\eta ^{\ast })K^{\ast
\mu }, \\
\mathcal{L}_{K^{\ast }N\Lambda } &=&-g_{K^{\ast }N\Lambda }\bar{N}\left( %
\rlap{$\slash$}K^{\ast }-\frac{\kappa _{K^{\ast }N\Lambda }}{2m_{N}}\sigma
_{\mu \nu }\partial ^{\nu }K^{\ast \mu }\right) \Lambda +\text{H.c.}~,
\end{eqnarray}%
where $\phi ^{\ast }$, $\eta ^{\ast }$, ${K}$, $K^{\ast }$, $N$, and $\Lambda
$ are the $\phi (2170)$, the $\eta (2225)$, the $K$, the t$K^{\ast }$ meson, the nucleon, and the $%
\Lambda $ fields, respectively. Here, by using the SU(3) flavor symmetry
relation \cite{Oh:2006hm,Oh:2006in}, the coupling constant $g_{KN\Lambda
}=-13.24$ can be determined \cite{Wan:2015gsl}. Moreover, we adopt coupling
constants $g_{K^{\ast }N\Lambda }=-4.26$ and $\kappa _{K^{\ast }N\Lambda
}=2.66$ calculated by the Nijmegen potential~\cite{Stoks:1999bz}. The value
of $g_{\phi ^{\ast }KK}$ and $g_{\eta ^{\ast }K^{\ast }K}$ can be determined
from the decay width $\Gamma _{\phi ^{\ast }\rightarrow KK}$ and $\Gamma
_{\eta ^{\ast }\rightarrow K^{\ast }K}$, respectively~\cite{Dong:2017rmg}. Accordingly, one gets $%
g_{\phi ^{\ast }KK}\simeq 2.81$ and $g_{\eta ^{\ast }K^{\ast }K}\simeq 1.39$
by taking $\Gamma _{\phi ^{\ast }\rightarrow KK}\simeq 81$ MeV and $\Gamma
_{\eta ^{\ast }\rightarrow K^{\ast }K}\simeq 79.2$ MeV, respectively.

\subsection{Amplitudes}

According to above Lagrangians, the scattering amplitude of the $%
K^{-}p\rightarrow \phi (2170)\Lambda $ and $K^{-}p\rightarrow \eta
(2225)\Lambda $ processes can be written as%
\begin{eqnarray}
i\mathcal{M}_{K^{-}p\rightarrow \phi (2170)\Lambda } &=&ig_{\phi ^{\ast
}KK}g_{KN\Lambda }F(q^{2})\epsilon _{\phi ^{\ast }}^{\mu }(k_{2})\bar{u}_N%
(p_{2})  \notag \\
&&\times \gamma _{5}\frac{1}{t-m_{K}^{2}}(k_{1\mu }+q_{\mu })u_\Lambda(p_{1}),
\label{AmpT1} \\
i\mathcal{M}_{K^{-}p\rightarrow \eta (2225)\Lambda } &=&g_{\eta ^{\ast
}K^{\ast }K}g_{K^{\ast }N\Lambda }F(q^{2})\bar{u}_N(p_{2})  \notag \\
&&\times \left( \gamma _{\nu }-\frac{\kappa _{K^{\ast }N\Lambda }}{2m_{N}}%
\gamma _{\nu }\rlap{$\slash$}q_{K^{\ast }}\right)  \notag \\
&&\times \frac{\mathcal{P}^{\mu \nu }}{t-m_{K^{\ast }}^{2}}(k_{2\mu
}+k_{1\mu })u_\Lambda(p_{1}),  \label{AmpT2}
\end{eqnarray}%
with%
\begin{equation}
\mathcal{P}^{\mu \nu }=i\left( g^{\mu \nu }+q_{K^{\ast }}^{\mu }q_{K^{\ast
}}^{\nu }/m_{K^{\ast }}^{2}\right) ,
\end{equation}%
where $\epsilon _{\phi ^{\ast }}^{\mu }$ is the polarization vector of the $%
\phi (2170)$ meson, and $\bar{u}_N$ or $u_\Lambda
$ is the Dirac spinor of nucleon or $%
\Lambda $ baryon. For the $t$-channel  exchange \cite{Liu:2008qx}, the
form factor $F(q^{2})=(\Lambda _{t}^{2}-m^{2})/(\Lambda _{t}^{2}-q^{2})$ is
adopted. Here, the $t=q^{2}=(k_{1}-k_{2})^{2}$ is the Mandelstam variables.
The cutoff $\Lambda _{t}$ in form factor is the only free parameter and will
be discussed in Sec. III.

\subsection{Reggeized $t$-channel}

Usually, the Regge trajectory{\ model is successful in analyzing hadron
production at high energies \cite%
{Wan:2015gsl,Wang:2015xwa,Haberzettl:2015exa,Wang:2015hfm,Ozaki:2010wp,Wang:2017qcw,Wang:2017plf}%
. The Reggeization can be done by replacing the $t$-channel propagator in
the Feynman amplitudes~(Eqs. \ref{AmpT1}$\ $and \ref{AmpT2}) }with the Regge
propagator as follows:
\begin{eqnarray}
\frac{1}{t-m_{K}^{2}} &\rightarrow &(\frac{s}{s_{scale}})^{\alpha _{K}(t)}%
\frac{\pi \alpha _{K}^{\prime }}{\Gamma \lbrack 1+\alpha _{K}(t)]\sin [\pi
\alpha _{K}(t)]}, \\
\frac{1}{t-m_{K^{\ast }}^{2}} &\rightarrow &(\frac{s}{s_{scale}})^{\alpha
_{K^{\ast }}(t)-1}\frac{\pi \alpha _{K^{\ast }}^{\prime }}{\Gamma \lbrack
\alpha _{K^{\ast }}(t)]\sin [\pi \alpha _{K^{\ast }}(t)]}.
\end{eqnarray}%
The scale factor $s_{scale}$ is fixed at 1 GeV. In addition, the Regge
trajectories $\alpha _{K}(t)$ and $\alpha _{K^{\ast }}(t)$ read as \cite%
{Ozaki:2010wp},%
\begin{equation}
\alpha _{K}(t)=0.70(t-m_{K}^{2}),\ \alpha _{K^{\ast
}}(t)=1+0.85(t-m_{K^{\ast }}^{2}).\quad \ \
\end{equation}%
It is noted that no additional parameter is introduced after the Reggeized
treatment applying.

\section{Numerical results}

With the preparation in the previous section, the cross section of the reactions $%
K^{-}p\rightarrow \phi (2170)\Lambda $ and $K^{-}p\rightarrow \eta
(2225)\Lambda $  can be calculated. The differential cross section
in the center of mass (c.m.) frame is written as
\begin{equation}
\frac{d\sigma }{d\cos \theta }=\frac{1}{32\pi s}\frac{\left\vert \vec{k}%
_{2}^{{~\mathrm{c.m.}}}\right\vert }{\left\vert \vec{k}_{1}^{{~\mathrm{c.m.}}%
}\right\vert }\left( \frac{1}{2}\sum\limits_{\lambda }\left\vert \mathcal{M}%
\right\vert ^{2}\right) ,
\end{equation}%
where $s=(k_{1}+p_{1})^{2}$, and $\theta $ denotes the angle of the outgoing
$\phi ^{\ast }/\eta ^{\ast }$ meson relative to $K$ beam direction in the
c.m. frame. $\vec{k}_{1}^{{~\mathrm{c.m.}}}$ and $\vec{k}_{2}^{{~\mathrm{c.m.%
}}}$ are the three-momenta of initial $K$ beam and final $\phi ^{\ast }/\eta
^{\ast }$, respectively.   One notes that both the $\phi(2170)$ and the $\eta(2225)$ have large widths. If we consider the width in the calculation, the thresholds will  not be fixed  but varied in an energy region.  As seen later, the best energy window to observe these two states is not close to the threshold. Hence, for simplification,  we do not consider the effect of the width on the thresholds.

\begin{figure}[h!]
\begin{center}
\includegraphics[scale=0.35]{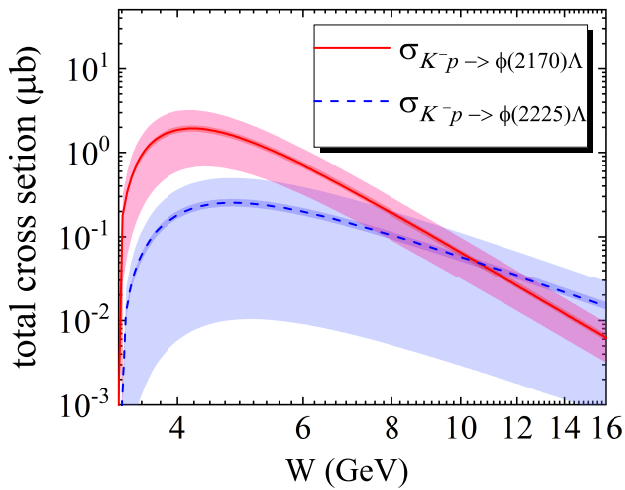}
\end{center}
\caption{(Color online) The energy dependence of the total cross section for
the productions of the $\protect\phi (2170)$ and the $\protect\eta (2225)$ through $t$
channel with cutoff $\Lambda _{t}=1.6\pm 0.5$ GeV. The Full (red) and dashed
(blue) lines are for the reactions $K^{-}p\rightarrow \protect\phi (2170)\Lambda $ and
$K^{-}p\rightarrow \protect\eta (2225)\Lambda $, respectively. The
bands stand for the error bar of cutoff $\Lambda _{t}.$}
\label{Fig:total}
\end{figure}

\begin{figure}[t]
\centering
\includegraphics[scale=0.4]{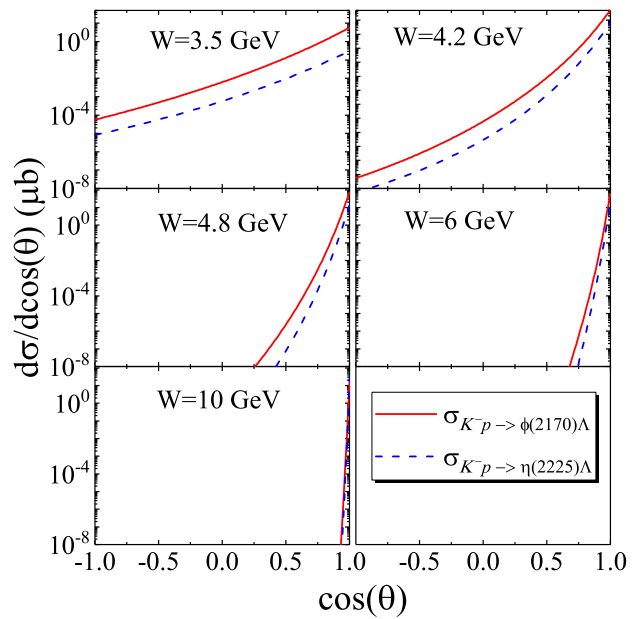}
\caption{(Color online) The differential cross section $d\protect\sigma %
/d\cos \protect\theta $ of the $\protect\phi (2170)$ and $\protect\eta %
(2225) $ production at different center-of-mass energies $W=3.5,4.2,4.8,6$
and $10$ GeV. The Full (red) and dashed (blue) lines are for the $%
K^{-}p\rightarrow \protect\phi (2170)\Lambda $ and $K^{-}p\rightarrow
\protect\eta (2225)\Lambda $ process, respectively. Here, one take the
cutoff $\Lambda _{t}=1.6$. }
\label{dcs}
\end{figure}

Since there does not exist the experimental data for the reactions $K^{-}p\rightarrow
\phi (2170)\Lambda $ and $K^{-}p\rightarrow \eta (2225)\Lambda $ ,
here we give the prediction of the cross section of these two reactions as
presented in Figs.~\ref{Fig:total}-\ref{dcs}. In these calculations, the
cutoff parameter in the form factor is the only free parameter. In Ref. \cite%
{Ozaki:2010wp}, for the Reggeized $t$ channel with $K$ or $%
K^{\ast }$ exchange, the experimental data can be reproduced well by taking
cutoff $\Lambda _{t}=1.55$ GeV. Moreover, in our previous work \cite%
{Wang:2018mjz}, the fitted value of free parameter $\Lambda _{t}=1.60\pm
0.03$ GeV was obtained for the reaction $\pi ^{-}p\rightarrow f_{1}(1420)n$  via $t$
channel in a Regge model, which is also a reasonable value for the reaction $%
K^{-}p\rightarrow f_{1}(1420)\Lambda $  through $K^{\ast }$ exchange.
Thus the cutoff $\Lambda _{t}\simeq 1.6$ GeV will be taken in the current work.

In Fig.~\ref{Fig:total} we present the total cross section of the reactions $%
K^{-}p\rightarrow \phi (2170)\Lambda $ and $K^{-}p\rightarrow \eta
(2225)\Lambda $  within the Regge model by taking $\Lambda
_{t}=1.6\pm 0.5$ GeV, respectively. It is found that the line shape of the total
cross section of the interaction $K^{-}p\rightarrow \phi (2170)\Lambda $  goes up
very rapidly and has a peak around $W=4.2$ GeV. The total cross section of the $%
\phi (2170)$ production in $K$ induced reaction can reach up about 1.9 $\mu $%
b at $W=4.2$ GeV, which indicates that the $W\simeq 4.2$ GeV is the best
energy window for searching for the $\phi (2170)$ via kaon induced reaction.

From Fig.~\ref{Fig:total} we also found that the total cross section of the $%
K^{-}p\rightarrow \eta (2225)\Lambda $ process
 goes up slowly and has a
bump at $4$ GeV $\lesssim $ $W\lesssim 6$ GeV because the $K^{\ast }$ exchange is dominant in the $\eta(2225)$ production while the $K$ exchange is dominant in the $\phi(2170)$ production~\cite%
{Ozaki:2010wp,Wang:2018mjz}.  At larger cutoff, the total cross section of $%
K^{-}p\rightarrow \eta (2225)\Lambda $ process is about  one order of magnitude lower than that for the $\phi(2170)$ production.  At smaller cutoff, the total cross section for the $\eta(2225)$ production decreases rapidly with the decrease of the cutoff due to the larger mass of the exchanged $K^*$ meson.

In Fig.~\ref{dcs}, we present the prediction of differential cross section
of the reactions $K^{-}p\rightarrow \phi (2170)\Lambda $ and $K^{-}p\rightarrow \eta
(2225)\Lambda $  within the Regge model by taking a cutoff $\Lambda
_{t}=1.6$ GeV. It can be seen that the differential cross sections of these two
reactions are very sensitive to the $\theta $ angle and show strong
forward-scattering enhancements especially at higher energies. Based on the
results, the measurement at forward angles is suggested, which can be used
to check the validity of the Reggeized treatment.

\section{Summary and discussion}

In this work, we study the reactions $K^{-}p\rightarrow \phi (2170)\Lambda $ and $%
K^{-}p\rightarrow \eta (2225)\Lambda $  with an effective Lagrangian
approach and the Regge trajectories model. The numerical results indicate
that the total cross section of the $\phi (2170)$ production
can reach an order of magnitude  of 1 $\mu $b at reasonable cutoffs, which means that it is feasible to
search for the $\phi (2170)$ by kaon beam.   Moreover, as we
expected, the differential cross sections of both reactions $K^{-}p\rightarrow \phi
(2170)\Lambda $ and $K^{-}p\rightarrow \eta (2225)\Lambda $  are
sensitive to the $\theta $ angle and gives a considerable contribution at
forward angles.

 The $\eta(2225)$ production is more dependent on the cutoff, and has a smaller cross section than the $\phi(2170)$.  The line shape of
total cross section shows the monotonically decreasing behavior, which is
different from the result with the Feynman model \cite{Ozaki:2010wp,Wang:2018mjz}%
.Usually, in the Regge
model the $K^{\ast }$ trajectory will be naturally decreasing. Besides,
there is a clear bump structure in the total cross section at $4$ GeV $%
\lesssim $ $W\lesssim 6$ GeV, which indicates that energies at $4-6$ GeV is
the best window for searching for the $\eta (2225) $ via reaction $K^{-}p\rightarrow
\eta (2225)\Lambda $ .

The high-precision data are expected at the facilities, such as the J-PARC, JLab, and COMPASS, which can provide good kaon beams. Our theoretical results provide valuable
information for these possible experiments of searching for the $\phi (2170)$ at these facilities.
The high precision data expected from the kaon induced interaction and other production mechnisms, such as photoproduction  at GlueX@CEBAF 12 GeV upgraded~\cite{Austregesilo:2018mno}, will provide a clearer picture of the
interaction mechanism of strange quarkoniums production.

\section{Acknowledgments}

This project is supported by the National Natural Science Foundation of
China under Grants No. 11705076 and No. 11675228. We acknowledge the Natural
Science Foundation of Gansu province under Grant No. 17JR5RA113. This work
is partly supported by the HongLiu Support Funds for Excellent Youth Talents
of Lanzhou University of Technology.

\end{document}